\documentclass{article}

\usepackage{arxiv}
\usepackage{algorithm}
\usepackage{algpseudocode}
\usepackage{amsfonts}
\usepackage{amsmath}
\usepackage{amssymb}
\usepackage{amsthm}
\usepackage{booktabs}
\usepackage{dsfont}
\usepackage{mathrsfs}
\usepackage{multirow}
\usepackage{pgfplots}
\usepgfplotslibrary{fillbetween}
\usepackage{tikz}
\usepackage{siunitx}
\usepackage{stmaryrd}
\usepackage{subcaption}
\usepackage{xcolor}
\usepackage{hyperref}
\usepackage{varwidth}
\usepackage[normalem]{ulem}
 \usepackage{setspace}

\usepgfplotslibrary{external} 
\tikzsetexternalprefix{ext/}

\makeatletter
\g@addto@macro\@floatboxreset\centering
\makeatother

\usetikzlibrary{calc,fadings,shapes.misc,3d,arrows,
                decorations,decorations.pathmorphing,
                decorations.text,shapes.geometric,
                decorations.markings,patterns,spy,positioning,
                arrows.meta,quotes}

\usepackage{tikz-3dplot}
\definecolor{myblue1}{RGB}{0,177,234}
\definecolor{myblue2}{RGB}{76,200,239}
\definecolor{myblue3}{RGB}{127,215,244}
\definecolor{myblue4}{RGB}{178,231,248}
\definecolor{mybluegray1}{RGB}{0,127,167}
\definecolor{mybluegray2}{RGB}{76,165,193}
\definecolor{mybluegray3}{RGB}{127,191,211}
\definecolor{mybluegray4}{RGB}{178,216,228}
\definecolor{mygray1}{RGB}{76,84,93}
\definecolor{mygray2}{RGB}{129,135,141}
\definecolor{mygray3}{RGB}{165,169,174}
\definecolor{mygray4}{RGB}{201,203,206}
\definecolor{myorange1}{RGB}{255,126,46}
\definecolor{myorange2}{RGB}{255,164,108}
\definecolor{myorange3}{RGB}{255,190,150}
\definecolor{myorange4}{RGB}{255,216,192}
\definecolor{mypurple1}{RGB}{89,89,171}
\definecolor{mypurple4}{RGB}{189,189,231}
\definecolor{back}{RGB}{200,197,190}

\newcommand{\eps}{\varepsilon}

\newcommand{\V}[1]{\textbf{#1}}
\newcommand{\GV}[1]{\boldsymbol{#1}}
\newcommand{\R}{\mathbb{R}}

\newcommand{\eg}{\textit{e.g.}\;}
\newcommand{\ie}{\textit{i.e.}\;}

\newcommand{\blue}[1]{#1}

\title{A supervised neural network for drag prediction of arbitrary 2D shapes in laminar flows at low Reynolds number}

\author{
    Jonathan Viquerat\thanks{Corresponding author}\\
    MINES Paristech , PSL - Research University\\
    CEMEF\\
    \texttt{jonathan.viquerat@mines-paristech.fr} \\
\And
    Elie Hachem\\
    MINES Paristech , PSL - Research University\\
    CEMEF\\
    \texttt{elie.hachem@mines-paristech.fr} \\
}

\begin{document}
\newgeometry{left=3cm,right=3cm,top=3cm,bottom=3cm}
\maketitle

\begin{abstract}
Despite the significant breakthrough of neural networks in the last few years, their spreading in the field of computational fluid dynamics is very recent, and many applications remain to explore. In this paper, we \blue{explore the drag prediction capabilities of convolutional neural networks} for \blue{laminar}, low-Reynolds number flows past arbitrary 2D shapes. A set of random shapes exhibiting a rich variety of geometrical features is built using B\'ezier curves. The efficient labelling of the shapes is provided using an immersed method to solve a unified Eulerian formulation of the Navier-Stokes equation. The network is then trained \blue{and optimized} on the obtained dataset, and its predictive efficiency assessed on several real-life shapes, including NACA airfoils.
\end{abstract}


\section{Introduction}

\begin{figure}[h!]
\centering

\hspace{-18pt} \includegraphics{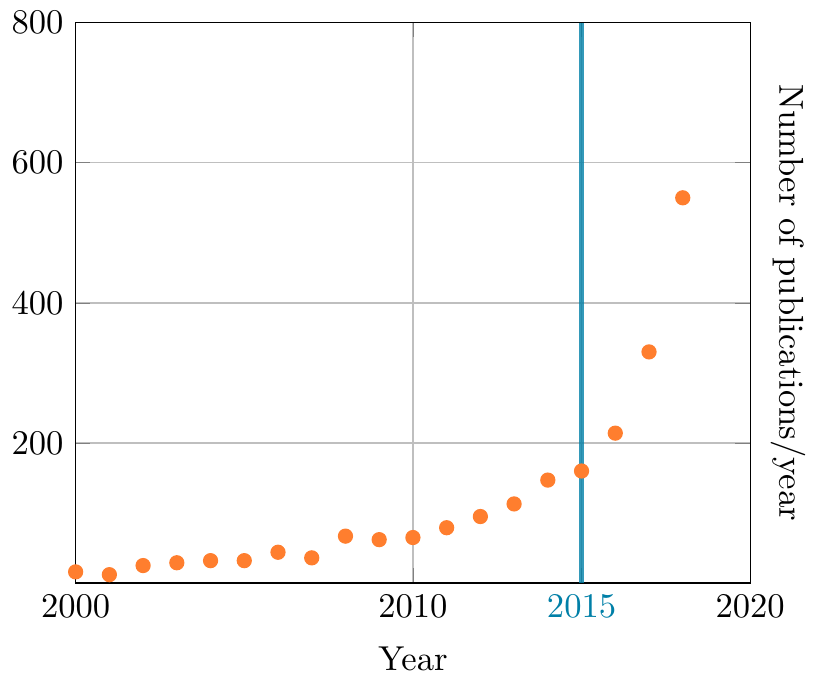}

\caption{\textbf{Number of publications matching keywords "machine learning", "neural networks" and "computational fluid dynamics"} in Google Scholar, between 2000 and 2018.}
\label{fig:papers}
\end{figure}

The recent successes of machine learning (ML), and more specifically neural networks (NN), have drawn increasing attention from the scientific community on the capabilities of such methods, and their possible applications to diverse research fields. In the computational fluid dynamics (CFD) field, the topic triggered a real enthusiasm from the year 2015, with a highly-increasing amount of related papers since (see figure \ref{fig:papers}). Despite this recent hype, much remains to be done before the possibilities and limits of such methods are well contoured.

In the recent years, neural networks have been used in very different ways in order to assist or improve CFD computations. Very often, a NN is used to replace one step of the resolution process, either to attain better performance or to extricate from a limited model and gain in generality. Examples for these applications are the replacement of the pressure projection step in Eulerian methods \cite{Tompson2016}, or the prediction of closure terms in RANS \cite{Ling2016} \cite{Tracey2015} or LES \cite{Beck2018} computations. Direct solving of Navier-Stokes equations can also be performed with NN using physics informed deep learning, where two networks are used concurrently \cite{Raissi2017} \cite{Raissi2018}. The first one is trained to predict the partial differential equation (PDE) solution, while the second one is used to incorporate constraints from the original PDE.

In other cases, a flow profile or a figure of merit (such as drag or lift) can be directly sought from a supervised network. \blue{Indeed, using a trained neural network as a surrogate model can be of particular interest in the context of optimization problems, or real-time decision processes.} In \cite{Zhang2017}, the authors focus on the prediction of lift for 2D airfoil profiles in different flow conditions and at different incidence angles. A key point of the latter contribution is the exploration of an original method of inputting flow conditions along with airfoil profile using an "artificial image", where free-space pixels around the airfoil are coloured depending on the value of the Mach number. In \cite{Guo2016}, convolutional neural networks (CNN) are trained to make visual predictions of the steady state flow around primitive shapes and real-life shapes, such as cars, using a signed distance map as input. In \cite{Miyanawala2017}, the authors explore the capabilities of a NN to map design parameters of \blue{geometrically primitive bluff bodies} to flow parameters, using a stochastic gradient descent method with momentum. \blue{One aspect of the current work is to extend the ideas of \cite{Miyanawala2017} to arbitrary 2D shapes.}

In this paper, we explore the predictive capabilities of a specific neural networks architecture at low Reynolds regime around 2D randomly generated shapes. In the first section, a brief overview of the general functioning of supervised NN is presented. Then, the dataset generation is addressed, along with the efficient resolution of the Navier-Stokes equations using an embedded mesh method. In the fourth section, a baseline convolutional network is introduced and optimized. Finally, the predictive capabilities of the network are explored on realistic configurations, such as geometrical shapes or airfoils. The base code used in this paper is available at \url{https://github.com/jviquerat/cnn_drag_prediction}.

\section{Neural networks}

Fundamentally, a neural network aims at approximating a function $f \colon V \to W$ that represents a complex and possibly implicit relation between two spaces of finite dimensions. In supervised learning, the network is exposed to a large set of couples $(\V{x} \in V, \V{y} \in W)$ which are known to \blue{fulfill} the relation $\V{y} = f(\V{x})$. For each couple, the network takes $\V{x}$ as an input, and outputs a prediction $\V{y}^*$. The error between $\V{y}^*$ and $\V{y}$ is computed, and fed back to the network, which internal parameters are adjusted accordingly \textit{via} an optimization algorithm. 

Classically, the tasks performed by neural networks are of two main kinds: (i) classification (\eg, analyzing handwritten text) or (ii) regression (\eg, predicting the lift of an airfoil from its shape). The goals of this paper fall under the second category. In the remaining of this section, we provide a brief description of the functioning of neural networks under supervised learning. Along the way, references to more thorough developments are also given.

\subsection{Artificial neurons and fully connected networks}

\blue{The basic unit of NN is the \emph{neuron}, to which an input vector $\V{x}$, associated to a set of weights $\V{w}$, is provided.} The neuron then computes the weighted sum $\V{w} \cdot \V{x} + b$, where $b$ is called the \emph{bias}, and applies the \emph{activation function} $\sigma$ to this sum. This is the output of the neuron, hereafter noted $z$. In the neuron, the weights and the bias represent the \emph{degrees of freedom} (\ie the parameters that can be adjusted to approximate the function $f$), while the activation function is a \emph{hyperparameter}, \ie it is part of the choices made during the network design.

\begin{figure}[h!]
\centering

\includegraphics{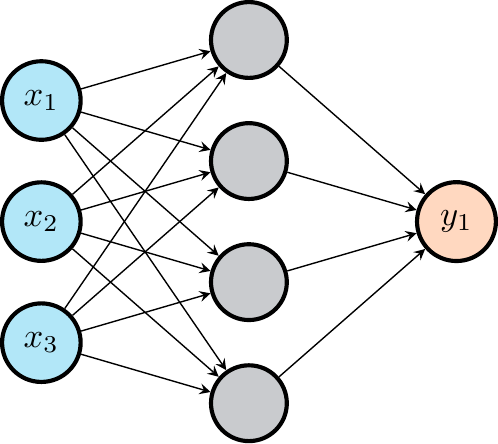}

\caption{\textbf{Simple example of neural network} with an input vector $\V{x} \in \R^3$, a hidden layer composed of 4 neurons, and an output layer composed of a single neuron. As a convention, input variables are drawn using a neuron representation. However, it must be kept in mind that the input layer is composed of neurons that simply map the identity.}
\label{fig:simple_network}
\end{figure}

In their simplest form, neural networks consist in several \emph{layers} of neurons connected together, as shown in the basic example of figure \ref{fig:simple_network}. This network is said to be \emph{fully connected} (FC), in the sense that each neuron of a layer is connected to all the neurons of the following layer. Hence, this network contains $3 \times 4$ weights and $4$ biases for the hidden layer, and $4$ weights and $1$ bias for the output layer, for a total of $21$ degrees of freedom. Along with the choice of the activation functions, the number and size of layers are also part of the hyperparameters. \blue{It should be noted that in such networks, (i) the neurons of the input layer simply map the identity, and (ii) for regression problems, a linear activation function is used for the neurons of the output layer.}

\subsection{Convolutional networks}

\begin{figure}
\centering

\begin{subfigure}{.45\textwidth}
	\includegraphics{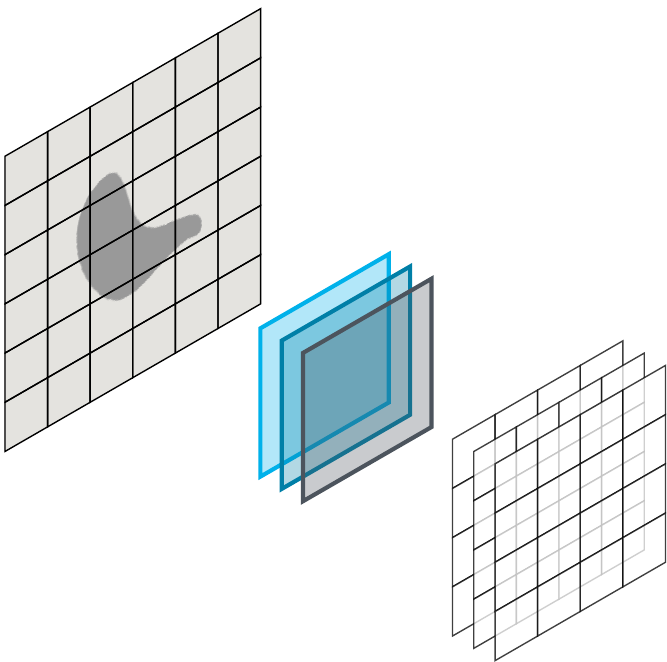}
	\caption{A convolutional layer holding three $3\times3$ kernels, thus producing three activation maps}
	\label{fig:cnn_kernel_1}
\end{subfigure} \qquad
\begin{subfigure}{.45\textwidth}
	\includegraphics{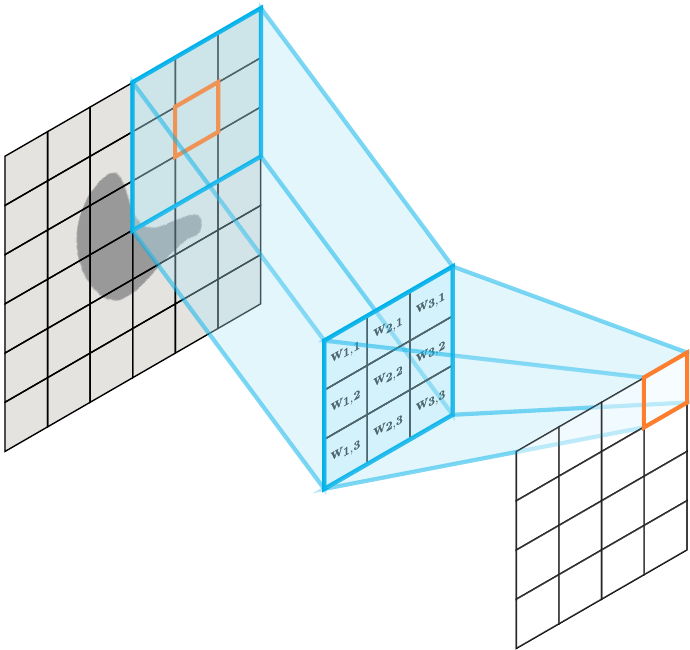}
	\caption{Convolution of a kernel with the input image}
	\label{fig:cnn_kernel_2}
\end{subfigure}

\caption{\textbf{Representation of a convolutional layer structure.} Left: structure of a CNN layer holding multiple convolutional kernels applied to the same input, each producing a specific activation map. Right: detail of a convolutional kernel applied to an input image and producing an activation map.}
\label{fig:cnn_kernel}
\end{figure}

When working with images as input (which will be the case in the following study), it is customary to exploit convolutional layers instead of FC ones. \blue{Each layer of a convolutional neural network (CNN) is composed of a set of convolution kernels that are used to extract spatial features from their input, as shown in figure \ref{fig:cnn_kernel_1}. During a forward pass, each kernel is convolved with its input to create an activation map, showing the response of the kernel at every spatial position (see figure \ref{fig:cnn_kernel_2}). The learnable weights of the network are the kernel parameters, such that, during the training, the network learns to extract spatial features that are meaningful to the current prediction problem.}

\begin{figure}
\centering

\includegraphics{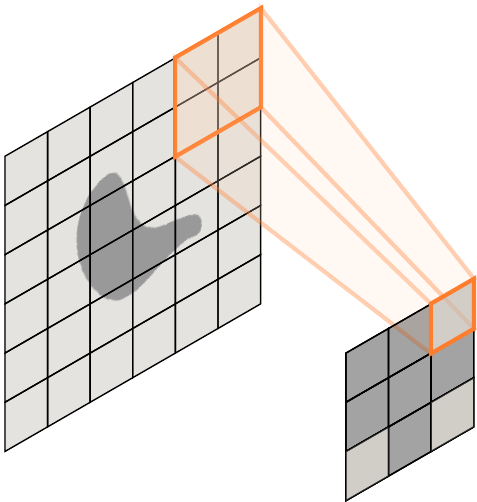}

\caption{\textbf{Representation of a max-pooling layer structure.} The output of the $2\times2$ max-pooling operation is the max value over its receptive field. Unlike convolutional layers, max-pooling layers usually use a stride equal to their size, meaning that there is no overlapping when the pooling operation is slided over the input image.}
\label{fig:cnn_pooling}
\end{figure}

\blue{Convolutional layers are often followed by \emph{pooling layers}, which role is to reduce the spatial size of the problem, which (i) helps to decrease the number of degrees of freedom in the network, and (ii) spreads the initial data throughout the successive convolutional layers. Today, max-pooling is used in a majority of cases, although other options are available, such as average-pooling, or L2-norm-pooling. A representation of a max-pooling layer is shown in figure \ref{fig:cnn_pooling}.}

\blue{CNNs present several advantages:}

\begin{itemize}
	\item \blue{\emph{Shared weights:} The sparse connection between the neurons of successive convolutional layers, and the use of the same weights for a given kernel over the whole input image greatly reduce the number of parameters to train in the network;}
	\item \blue{\emph{Efficient spatial feature extraction:} The use of convolutional kernels allows an efficient detection of spatial features at each level of the convolutional network. As successive layers of convolution and pooling are applied, more and more complex spatial features are extracted;}
	\item \blue{\emph{Translational invariance:} A particular strength of convolutional layers is that they are able to detect specific features (textures, edges, shapes) with the same efficiency in different locations of an input picture. This property is called \emph{translational equivariance}, and implies that the position at which a feature is detected modifies the feature map obtained from the considered kernel. The conjunction of convolutional layers with pooling layers helps achieve \emph{translational invariance}, in the sense that the position at which a feature was previously detected will matter less and less as the spatial dimension is reduced by pooling operations.}
\end{itemize}

\blue{Although it is not systematic (as for fully convolutional networks \cite{Sofka2017}), CNNs can be terminated with several fully connected layers}, followed by an output layer, which size is determined by that of the sought quantity of interest. These considerations are discussed in details in a large variety of books and articles. For the sake of brevity, we refer the reader to \cite{Goodfellow2017} and the references therein for complementary informations.

\subsection{Technicalities}

This section briefly addresses several key points of neural networks that will be used in the remaining of the paper. Again, this barely represents an overview of these questions, and the reader is once again referred to \cite{Goodfellow2017} for a thorough discussion of each of them.

\subsubsection{Data pre-processing}
\label{section:preprocessing}

The pre-processing of data fed to neural networks is crucial, in the sense that it may significantly influence its ability to learn. In the following, inputs are composed of images of $p \times p$ pixels with one channel (black and white image), the pixel values ranging from 0 to 255. During the pre-processing step, these images are downscaled to $n \times n$ pixels (with $n \leq p$), and the pixel values are rescaled between 0 and 1. The reason behind this normalization is that feeding large (and inhomogeneous) values to a network can prevent the gradient descent of the back-propagation method to converge \cite{Chollet2018}.

Most often, the input dataset is split in three subsets: (i) a \emph{training set}, on which the learning will be performed, (ii) a \emph{validation set}, which is used to monitor the network accuracy periodically during training, and (iii) a \emph{test set}, on which the final performance of the network is assessed. The validation and test sets must not overlap with the training set, nor between them.

\blue{Although it does not fully apply to the current problem, a remark must be made about the possibility to use data augmentation. In cases where a small transformation of an input image (homothetic transformation, rotation, translation) should not modify its associated label (as for most classification tasks, for example), the available dataset can be made artificially larger by adding transformed input images to it, associated to their original label. In our case, as any transformation of an input shape would modify its resulting drag and lift (except up-down flip of the shape that would only change the sign of the lift), this method will not be exploited here.}

\subsubsection{Activation functions}

Activation functions are used to obtain a \emph{non-linear} mapping between the input and the output spaces. They are commonly chosen layer-wise. For classification cases, it is common to use sigmoids or hyperbolic tangents in the hidden layers, as they will stretch the input space around a central point, thus helping to separate elements from different classes. For regression cases, the \emph{rectified linear units} activation function (also called \emph{ReLU}) has proven to be a robust choice.

\subsubsection{Loss function and backpropagation}

The learning process in neural networks consists in adjusting all the biases and weights of the network in order to reduce the value of a well-chosen loss function. For regression cases, it is common to choose the mean squared error. With the loss function at hand, the optimization of the weights and biases is performed with a \emph{stochastic gradient descent} (SGD). \blue{This algorithm is based on the chain rule, and is at the core of the learning process, since it allows to compute the contribution of each degree of freedom of the network to the loss value.}


\subsubsection{Network size, overfitting and regularization}
\label{section:overfitting}

During the training step, the network is exposed multiple times to the same input data. A full iteration over all the input samples is called an \emph{epoch}, and it is not rare for advanced networks to be trained for hundreds or thousands of epochs. Inside each epoch, the network is exposed to random batches of input data, a step of SGD being performed after each batch.

\blue{The size of a neural network (number of layers and number of neurons in each layer) is a central question when designing a network. For a given task, a too small network will not be able to grasp the complexity of the implicit function to map. At the opposite, a network with too many degrees of freedom will end up overfitting, \ie it will fit so closely to the training set that it will be unable to generalize to new data.}

Several methods are available to limit overfitting. The first one consists in gathering more data, although this is often either impossible or expensive. A second option is to reduce the size of the network, in order to better balance the number of free parameters with the size of available input data. However, this may also lead to a loss in the generalization capabilities. The last option consists in using \emph{regularization}, which can be done in two ways:

\begin{enumerate}
	\item A penalisation term proportional to the squared weights ($l^2$ regularization) or their absolute value ($l^1$ regularization) can be applied to the loss function. This will globally constrain the weights to be smaller, which will favor the emergence of simpler features over complex (and specific) ones;
	\item A \emph{dropout} layer can also be applied between two hidden layers: this consists in randomly setting to zero a fraction of the information passing from one layer to the next. The goal is to introduce some random noise in the information travelling through the network in order to prevent fortuitous patterns to be learned.
\end{enumerate}

\blue{Overfitting can also be tempered by using \emph{early stopping} \cite{Caruana2001}, which most often consists in monitoring the loss for the validation set during training, and to stop training when this loss stops improving. In practice, an additional parameter can be set that delays early stopping of a given amount of epochs, to avoid premature stopping of the learning.}

\subsubsection{Neural network \blue{implementation}}
\label{section:nn_imp}

The amount of ready-to-use neural networks libraries has exploded in the recent years, most of them exploiting C++ or Python. For supervised learning, they usually include a wide range of choices regarding layer types, activation functions, losses, optimizers and so on. In this paper, we chose to use Keras \cite{Chollet2018} (with Tensorflow backend) for its high level of abstraction and the ease of use provided by the Python language.

\begin{figure}

\includegraphics{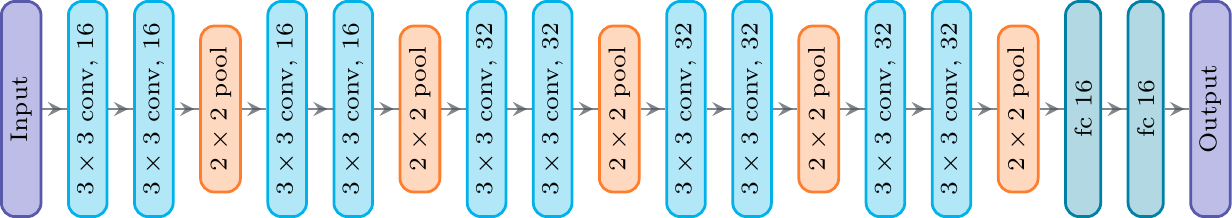}

\caption{\textbf{Baseline drag prediction CNN.} This network is based on a pattern made of two convolutional layers (in light blue) followed by a max-pooling layer (in orange). The pattern is repeated five times, the image size being divided by two each time. The last max-pooling layer is followed by two fully-connected layers (in dark blue). It is terminated with a fully-connected layer of size $1$ that outputs the predicted drag through a linear activation function.}
\label{fig:convnet_drag}
\end{figure}

\blue{In the remaining of this paper, we consider the general convolutional network architecture presented in figure \ref{fig:convnet_drag}, which was implemented using the basic Keras layers. In this network, a convolution/pooling pattern is repeated several times, with a variable number of filters for each layer (see figure \ref{fig:convnet_drag}). Here, the number of convolution layers in the base pattern is set to 2, with 8 filters in the initial layer. The network is terminated with 2 dense layers of size 16. The last layer is used to output the predicted drag and uses a linear activation function, while all the other layers of the network use ReLU activations. In this paper, network training is performed on a Tesla V100 GPU card, without pre-training (the network is systematically trained from scratch).}

\section{Dataset generation}

This section describes the generation of the dataset used in the remaining of the paper. First, we describe the steps to generate arbitrary shapes by means of connected Bezier curves. Then, the solving of the Navier-Stokes equations with an immersed method is presented. Finally, details about the dataset are given.

\subsection{Random shape generation}

\begin{figure}
\centering

\begin{subfigure}{.45\textwidth}
	\centering
	\includegraphics{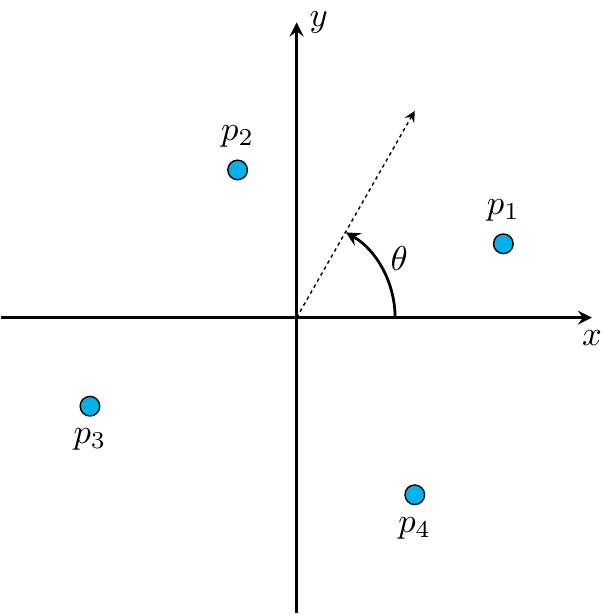}
	\caption{Draw $n_s$ random points, translate them around $(0,0)$ and sort them by ascending trigonometric angle}
	\label{fig:shape_generation_1}
\end{subfigure} \qquad
\begin{subfigure}{.45\textwidth}
	\centering
	\includegraphics{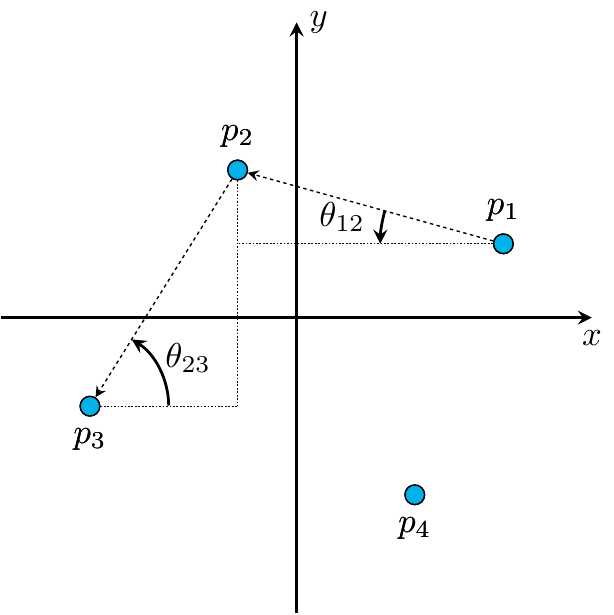}
	\caption{Compute angles between random points, and compute an average angle around each point $\theta^*_i$}
	\label{fig:shape_generation_2}
\end{subfigure}

\medskip
\medskip

\begin{subfigure}{.45\textwidth}
	\centering
	\includegraphics{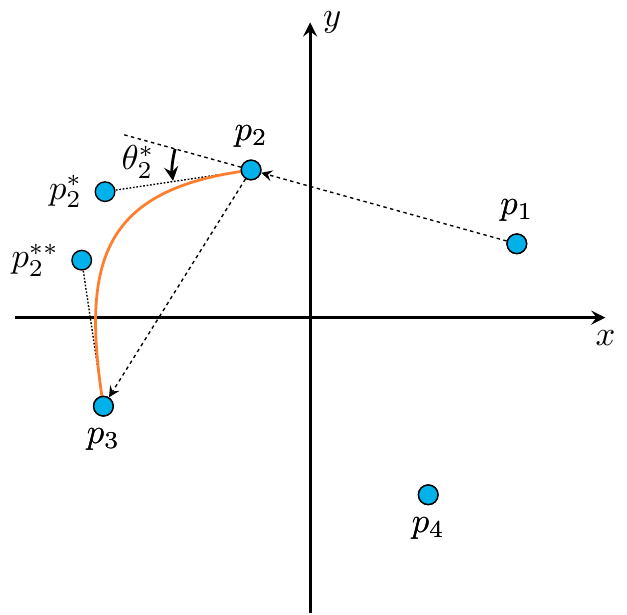}
	\caption{Compute control points coordinates from averaged angles and generate cubic B\'ezier curve}
	\label{fig:shape_generation_3}
\end{subfigure} \qquad
\begin{subfigure}{.45\textwidth}
	\centering
	\includegraphics{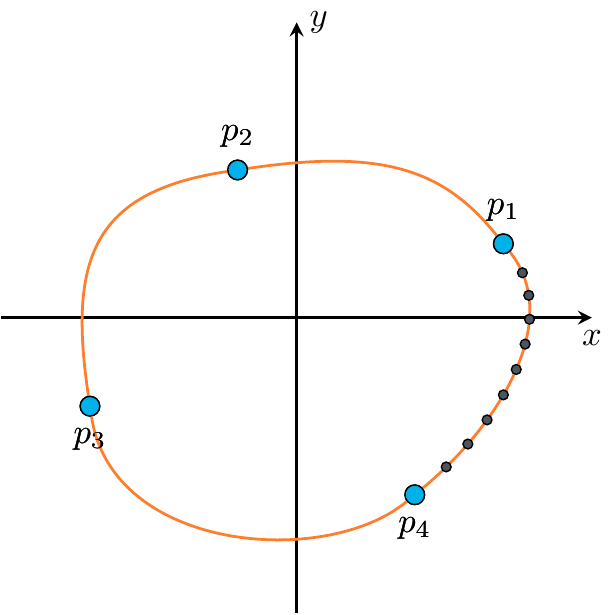}
	\caption{Sample all B\'ezier lines and export for mesh immersion}
	\label{fig:shape_generation_4}
\end{subfigure}

\caption{\textbf{Random shape generation with cubic B\'ezier curves}. }
\label{fig:shape_generation}
\end{figure}

The first step of the random shape generation consists in drawing $n_s$ random points in $\left[ 0, 1 \right]^2$, that are then translated so their center of mass is in $(0,0)$. The points are then sorted by ascending trigonometric angle (see figure \ref{fig:shape_generation_1}). The angles between consecutive random points are then computed, and an average is computed around each point (see figure \ref{fig:shape_generation_2}):

\begin{equation*}
	\theta^*_i = \alpha \theta_{i-1,i} + (1 - \alpha) \theta_{i,i+1},
\end{equation*}

\noindent with $\alpha \in \left[0,1\right]$. Averaging angles in such way will help smooth the final obtained shape, \blue{and in the remaining of this paper, $\alpha = 0.5$}. In the next step, a third order B\'ezier curve is drawn between each point, using the averaged angles $\theta^*_i$. Cubic B\'ezier curves are defined by four points: the first and last points, $p_i$ and $p_{i+1}$, are part of the curve, while the second and third ones, $p^*_i$ and $p^{**}_i$, are control points that define the tangent of the curve at $p_i$ and $p_{i+1}$. In our case, the tangents at $p_i$ and $p_{i+1}$ are determined respectively by $\theta^*_i$ and $\theta^*_{i+1}$ (see figure \ref{fig:shape_generation_3}). In a final step, all the Bezier curves are sampled, and a closed loop is exported to be used as an immersed mesh in a Navier-Stokes numerical simulation (figure \ref{fig:shape_generation_4}).

\begin{figure}
\centering

\includegraphics{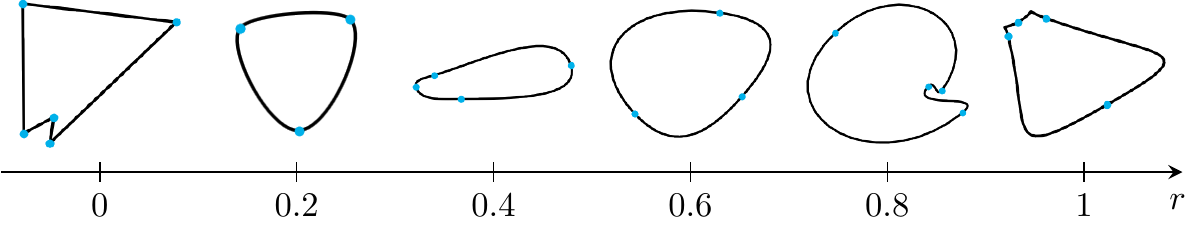}

\caption{\textbf{Random shape examples depending on their $r$ value,} ranging from 0 to 1. The random points are shown in blue, and their number $n_s$ ranges from 3 to 5, although it is possible to use more. For $r=0$, one sees the sharp features of the curve on the B\'ezier points. For intermediate values, smooth curves are obtained. Finally, for values close to 1, sharp features start to appear around the control points (not shown here).}
\label{fig:shape_examples}
\end{figure}

The sharpness of the curve features is handled with a positive parameter $r$ that controls the distances $[p_i p^*_i]$ and $[p_{i+1} p^{**}_i]$. For $r=0$, $p^{*}_i$ and  $p^{**}_i$ respectively coincide with $p_i$ and $p_{i+1}$, and the curve presents sharp angles at the control points. Intermediate values of $r$ produce smooth curves, with maximal smoothness for $r=0.5$. When increasing further toward $r=1$, sharp features start to appear near the crossing of the initial and final curve tagents. Finally, for $r > 1$, tangled cases start to appear. A variety of shapes obtained with different values of $r$ can be found in figure \ref{fig:shape_examples}. In the following, we restrict $r$ to the interval $\left[0, 1\right]$ to avoid tangled shapes.

\subsection{Navier-Stokes equations}
\label{section:NS}

\begin{figure}
\centering

\setlength{\fboxsep}{0pt}%
\setlength{\fboxrule}{1pt}%

\def\scale{1}

\begin{subfigure}{.45\textwidth}
	\centering
	\fbox{\includegraphics[width=\scale\linewidth]{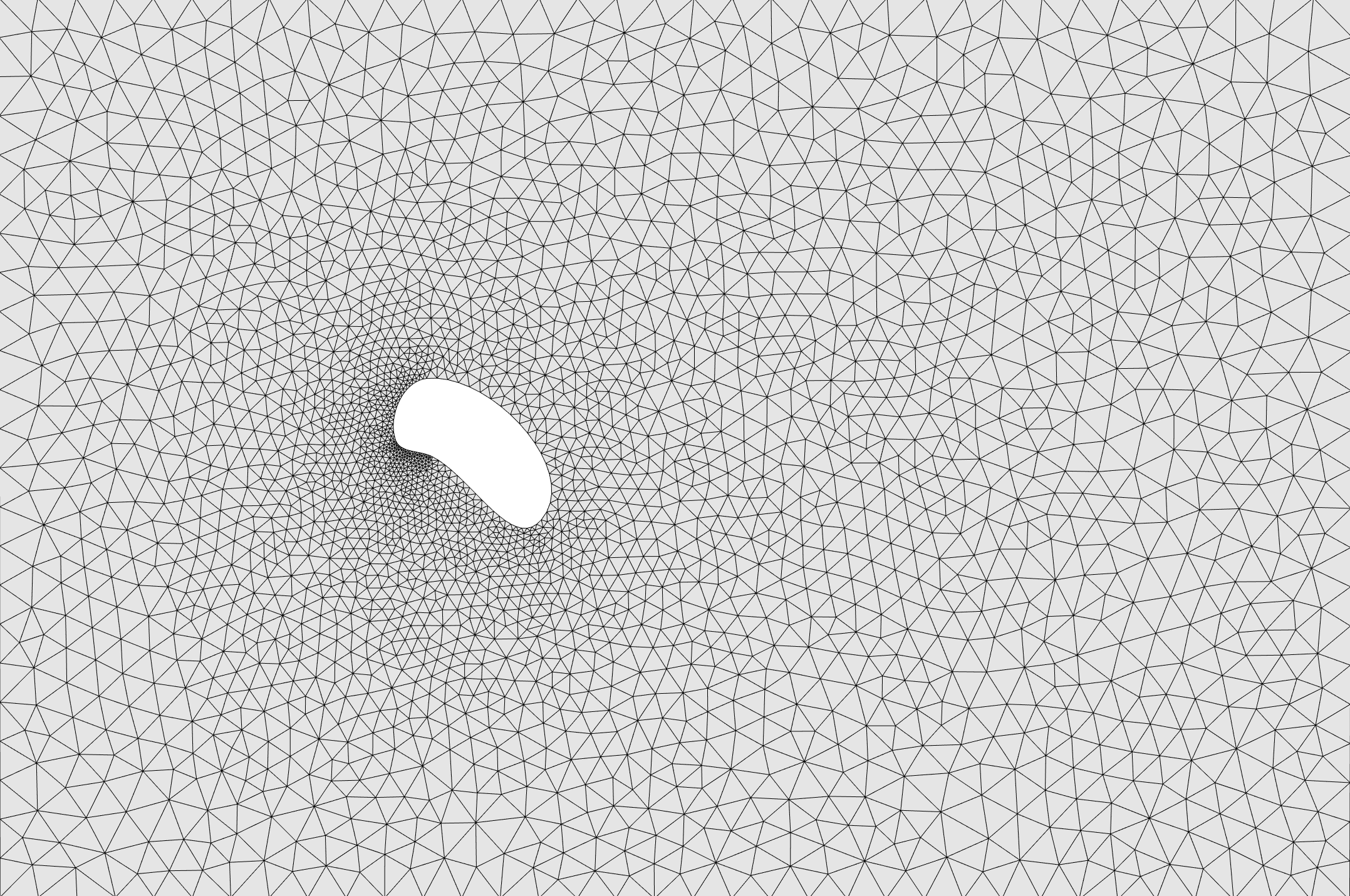}} 
	\caption{Body-fitted case}
	\label{fig:mesh_shape_bf}
\end{subfigure} \quad
\begin{subfigure}{.45\textwidth}
	\centering
	\fbox{\includegraphics[width=\scale\linewidth]{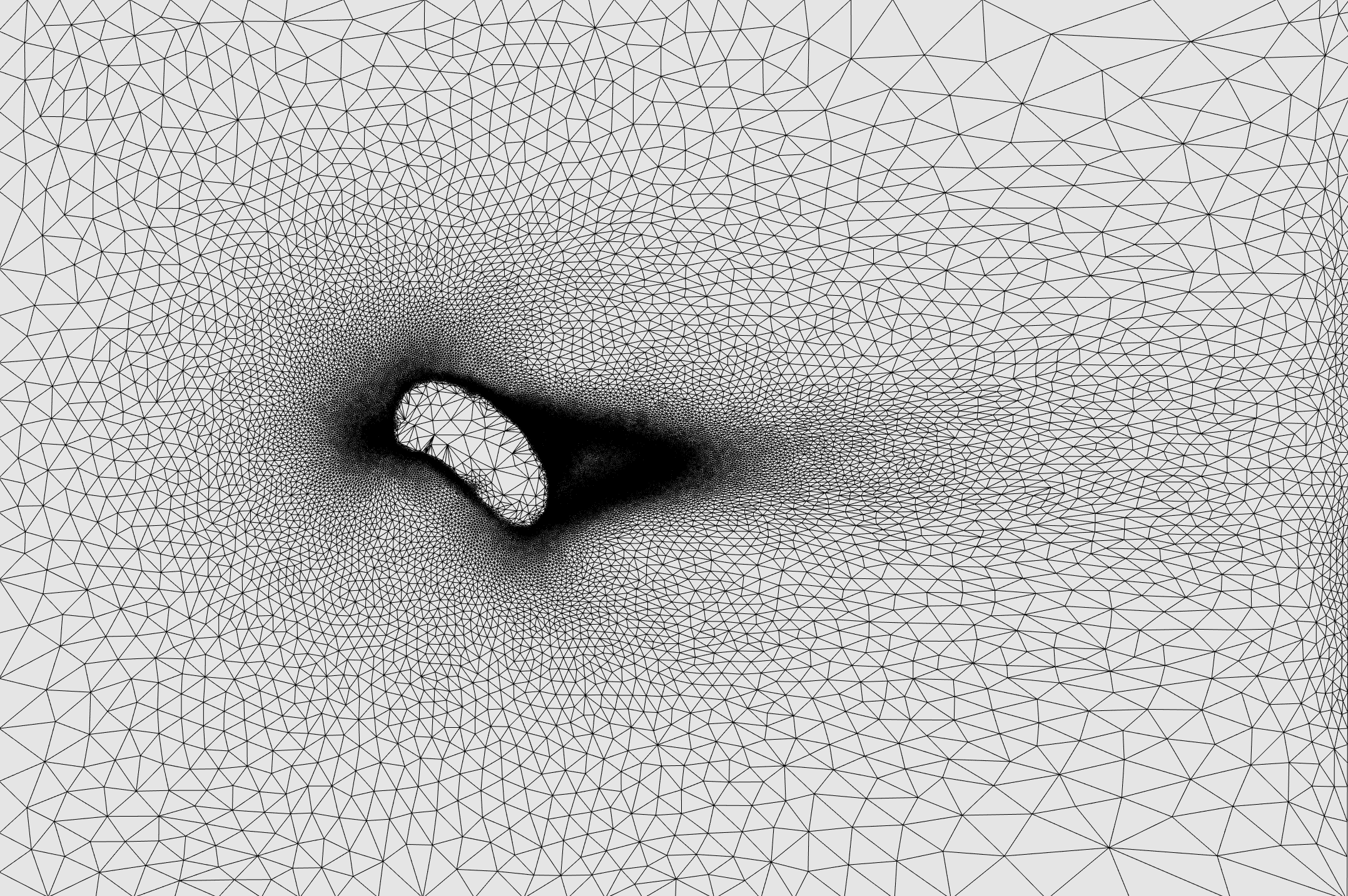}}
	\caption{Immersed case}
	\label{fig:mesh_shape_immersed}
\end{subfigure}

\caption{\textbf{The same shape meshed either in the body-fitted case (left) or with an immersed method (right).} In the immersed case, a regular remeshing is applied to better capture the velocity gradients.}
\label{fig:mesh_shape}
\end{figure}

The flow motion of incompressible Newtonian fluids is described by the Navier-Stokes (NS) equations: 

\begin{equation} \label{eq:ns_equation1}
	\left\{
	\begin{aligned}
		\rho\ (\partial_{t} \V{v} + \V{v} \cdot \nabla \V{v}) -\nabla \cdot \left( 2 \eta \GV{\epsilon}(\V{v}) - p \V{I} \right)  & = \V{f}, \\
		\nabla \cdot \V{v} &= 0,
	\end{aligned}
	\right.
\end{equation}

\noindent where $t \in [0,T]$ is the time, $\V{v}(x,t)$ the velocity, $p(x,t)$ the pressure, $\rho$ the fluid density, $\eta$ the dynamic viscosity, $\GV{\epsilon}$ the strain rate tensor and $\V{I}$ the identity tensor. Classically, the solving of NS equations around solid obstacles relies on body-fitted methods, where the mesh boundary follows the geometry of the obstacle. These methods require the generation of a full mesh (domain and obstacle) for each computation, which can be both time- and memory-consuming (see figure \ref{fig:mesh_shape_bf}). To overcome this issue, immersed methods based on a unified Eulerian formulation were introduced that propose to immerge a boundary mesh of the obstacle in a background mesh (see figure \ref{fig:mesh_shape_immersed}). The outline of this method is sketched in the next section.

\subsection{Interface description}
\label{section:immersed}

The formulation presented in this section is based on the introduction of an extra stress in the momentum equation of (\ref{eq:ns_equation1}). This extra stress is related to the appropriate deformation tensor in the solid domain and acts as a Lagrange multiplier to enforce that the deformation be zero in the solid. 

In the fluid-structure interaction field, monolithic approaches impose the use of an appropriate constitutive equation describing both the fluid and the solid domain. This offers a great flexibility to deal with different shapes in similar configurations without having to systematically re-mesh the whole domain. To do so, one starts by computing the signed distance function (level set) of the given geometry to each node of the background mesh: 

\begin{equation}
	\alpha(\V{x}) = \pm d\left( \V{x}, \Gamma_{\text{im}} \right), \forall \V{x} \in \Omega.
\end{equation}

\noindent Using this function, the fluid-solid interface $\Gamma_{\text{im}}$ is easily identified as the zero iso-value of function $\alpha$:

\begin{equation} \label{eq:levelset30}
	\Gamma_{\text{im}} = \left\{ \V{x} \in \Omega, \alpha(\V{x}) = 0 \right\}.
\end{equation}

\noindent In this paper, the following sign convention is used: $\alpha \geq 0$ inside the solid domain defined by the interface $\Gamma_{\rm im}$, and $\alpha \leq 0$  outside this domain. Further details about the algorithm used to compute the distance are available in \cite{Bruchon2009}. It is also possible to use functions  smoother than $d\left(\V{x},\Gamma_{\text{im}}\right)$ away from $\Gamma_{\text{im}}$ (see for example \cite{Codina2002}).

As explained above, the signed distance function is used to localize the interface of the immersed structure, but it is also used to initialize the desirable properties on both sides of the latter. Indeed, for the elements crossed by the level-set functions, fluid-solid mixtures are used to determine the element effective properties. To do so, a Heaviside function $H(\alpha)$ is defined as follows:

\begin{equation} \label{eq:heavyside31}
	H(\alpha) = \left\{
	\begin{aligned}
		1 & \text{ if}\ \alpha > 0,\\
		0 & \text{ if}\ \alpha < 0.
	\end{aligned}
	\right.
\end{equation}

\noindent The Heaviside function can be smoothed to obtain a better continuity at the interface \cite{Pijl2005} using the following expression:

\begin{equation} \label{eq:heavyside32}
	H_\varepsilon(\alpha) = 
	\begin{cases}
		\, 1 																		& \text{if } \alpha > \eps, \\
		\displaystyle \frac{1}{2} \left(1 + \frac{\alpha}{\eps} + \frac{1}{\pi} \sin \left( \frac{\pi \alpha}{\eps} \right) \right) 	& \text{if} \left|\alpha\right| \leq \eps,\\
		\, 0 																		& \text{if } \alpha < -\eps,
	\end{cases}
\end{equation}

\noindent where $\eps$ is a small parameter such that $\eps=O(h_{\text{im}})$, known as the interface thickness, and $h_{\text{im}}$ is the mesh size in the normal direction to the interface.

\subsection{Modified governing equations}
\label{section:equations}

Now that each system is expressed in an eulerian framework, we solve one global NS set of equations using the geometrical representation given by $H(\alpha)$ as follows:

\begin{equation} \label{eq:ns_equation1}
	\left\{
	\begin{aligned}
		\rho^* (\partial_{t} \V{v} + \V{v} \cdot \nabla \V{v}) -\nabla \cdot \left( 2 \eta \GV{\epsilon}(\V{v}) + \GV{\tau} - p \V{I} \right)  & = \V{f}, \\
		\nabla \cdot \V{v} &= 0,
	\end{aligned}
	\right.
\end{equation}

\noindent where we have introduced the following mixed quantities:

\begin{equation*}
	\begin{aligned}
		\GV{\tau} & = H(\alpha) \GV{\tau}_{\text{s}},\\
		\rho^* & = H(\alpha) \rho_{\text{s}} + (1-H(\alpha)) \rho_{\text{f}},
	\end{aligned}
\end{equation*}

\noindent the subscripts $f$ and $s$ referring respectively to the fluid and to the solid. In the latter equalities, $\GV{\tau}_{\text{s}}$ acts as a Lagrange multiplier that yields $\GV{\epsilon}(\V{v}) = \V{0}$ in the solid \cite{Hachem2013} \cite{Jannoun2015}.

Eventually, the modified equations (\ref{eq:ns_equation1}) are cast into a stabilized finite element formulation, and solved using a variational multi-scale (VMS) solver (the reader is invited to refer to \cite{Hachem2013} for more details).

\subsection{Dataset}

\begin{figure}
\centering

\setlength{\fboxsep}{0pt}%
\setlength{\fboxrule}{1pt}%

\def\scale{0.738}

\begin{subfigure}{.3\textwidth}
	\centering
	\fbox{\includegraphics[width=\linewidth]{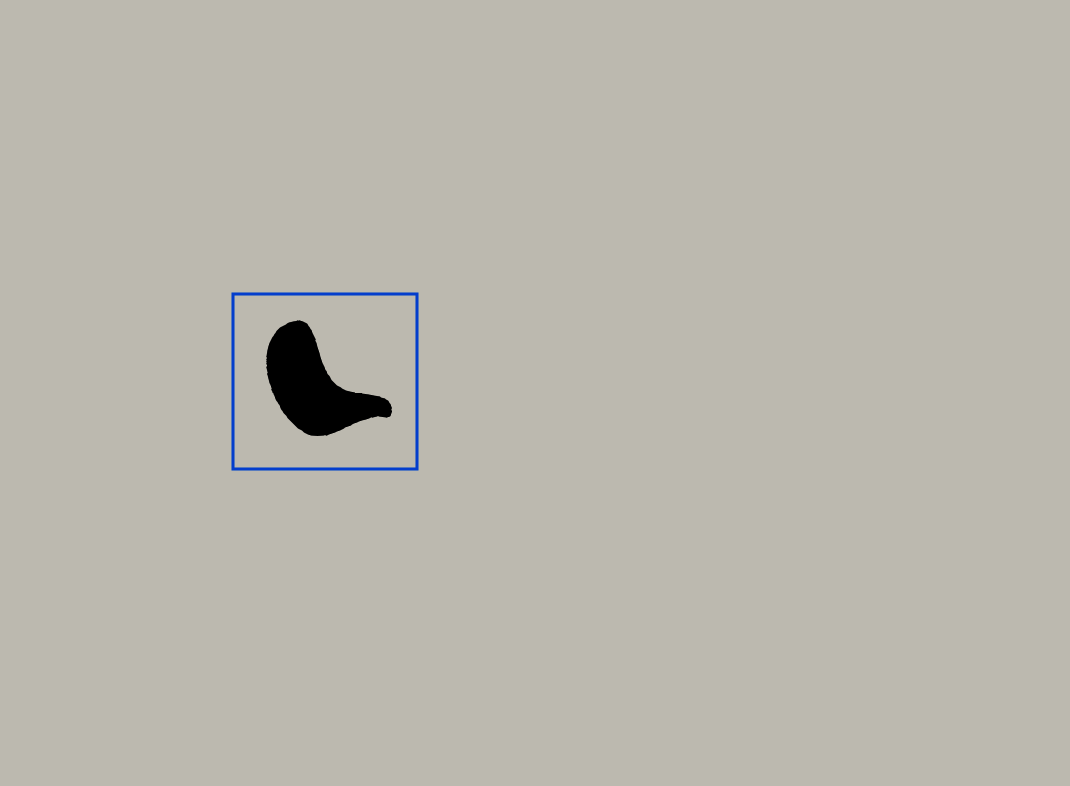}} 
	\caption{Computational domain}
\end{subfigure} \qquad
\begin{subfigure}{.3\textwidth}
	\centering
	\fbox{\includegraphics[width=\scale\linewidth]{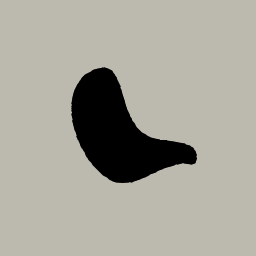}}
	\caption{Network input}
\end{subfigure}

\caption{\textbf{Computational domain and network input for a dataset element.}. The shape is shown in its computational domain (left), the blue frame indicating the actual subset of the image provided to the network. \blue{The subset size is chosen to be slightly larger than the maximal possible extent of the shape (which is known from the construction process), to avoid the shape touching the border of the frame.} A zoom of this subset is shown (right).}
\label{fig:dataset_example}
\end{figure}

\begin{figure}
\centering

\begin{subfigure}[t]{.45\textwidth}
	\hspace{-18pt} \includegraphics{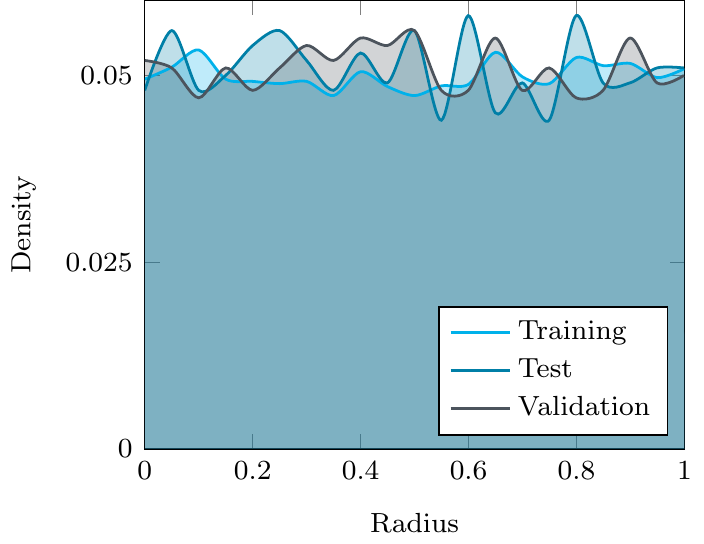}
	\caption{Radius density function}
\end{subfigure} \quad
\begin{subfigure}[t]{.45\textwidth}
	\hspace{-18pt} \includegraphics{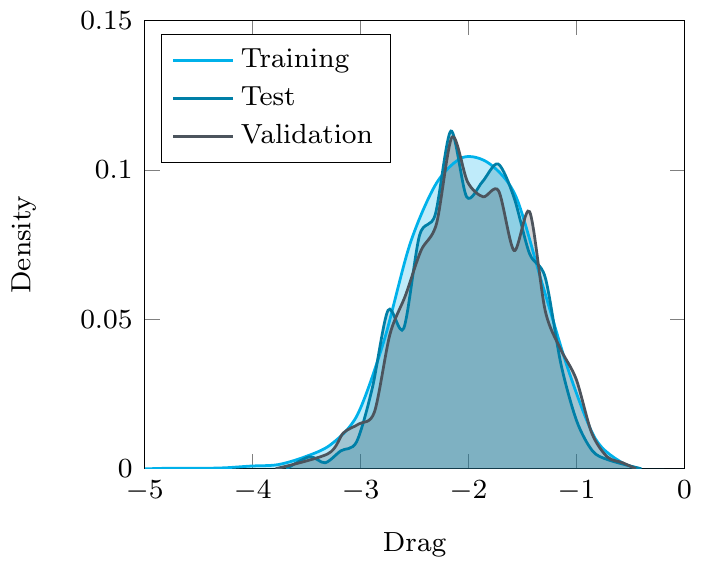}
	\caption{Drag density function}
\end{subfigure}

\caption{\textbf{Normalized density functions for the repartition of radius and drag values in the different data subsets.} The radius is drawn from a uniform probability density function on $\left[0,1\right]$. This results in a Gaussian-like drag distribution over the different subsets. In both cases, the distributions over the test and validation subsets are comparable.}
\label{fig:statistics}
\end{figure}

The dataset (DS) is composed of 12,000 shape images of size $128\times128$, along with their steady-state drag value at $Re=10$ (see figure \ref{fig:dataset_example}). To ensure a large diversity of shapes, $n_s$ is evenly distributed in $\left[3, 5\right]$, and $r$ in $\left[0, 1\right]$. In the following, the DS is systematically divided into three sets: 9600 shapes for the training set, 1200 shapes for the validation set, and 1200 shapes for the test set. During the dataset generation, a minimal distance between two nodes was prescribed, so that no shape can be smaller than size 0.1. No data augmentation was used, although up-down flip could have been used here, as stated in section \ref{section:preprocessing}. \blue{Statistics about the radius and drag repartitions over these subsets are shown in figure \ref{fig:statistics}. Radii values are comparably distributed over the different subsets, and this quasi-uniform distribution results in comparable Gaussian-like distribution of drags over the subsets.}

\begin{figure}
\centering

\setlength{\fboxsep}{0pt}%
\setlength{\fboxrule}{1pt}%

\begin{subfigure}[b]{.45\textwidth}
	\hspace{-18pt} \includegraphics{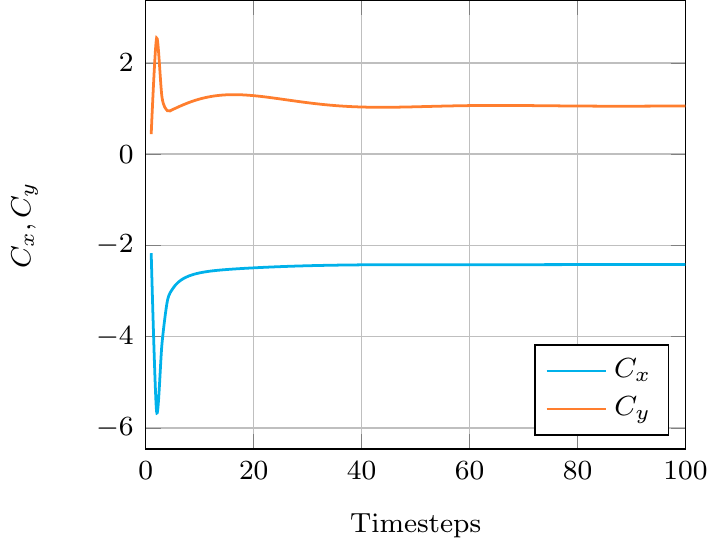}
	\caption{Evolution of the drag and lift coefficients during a CFD computation}
\end{subfigure} \quad
\begin{subfigure}[b]{.45\textwidth}
	\centering
	\fbox{\includegraphics[width=.8\linewidth]{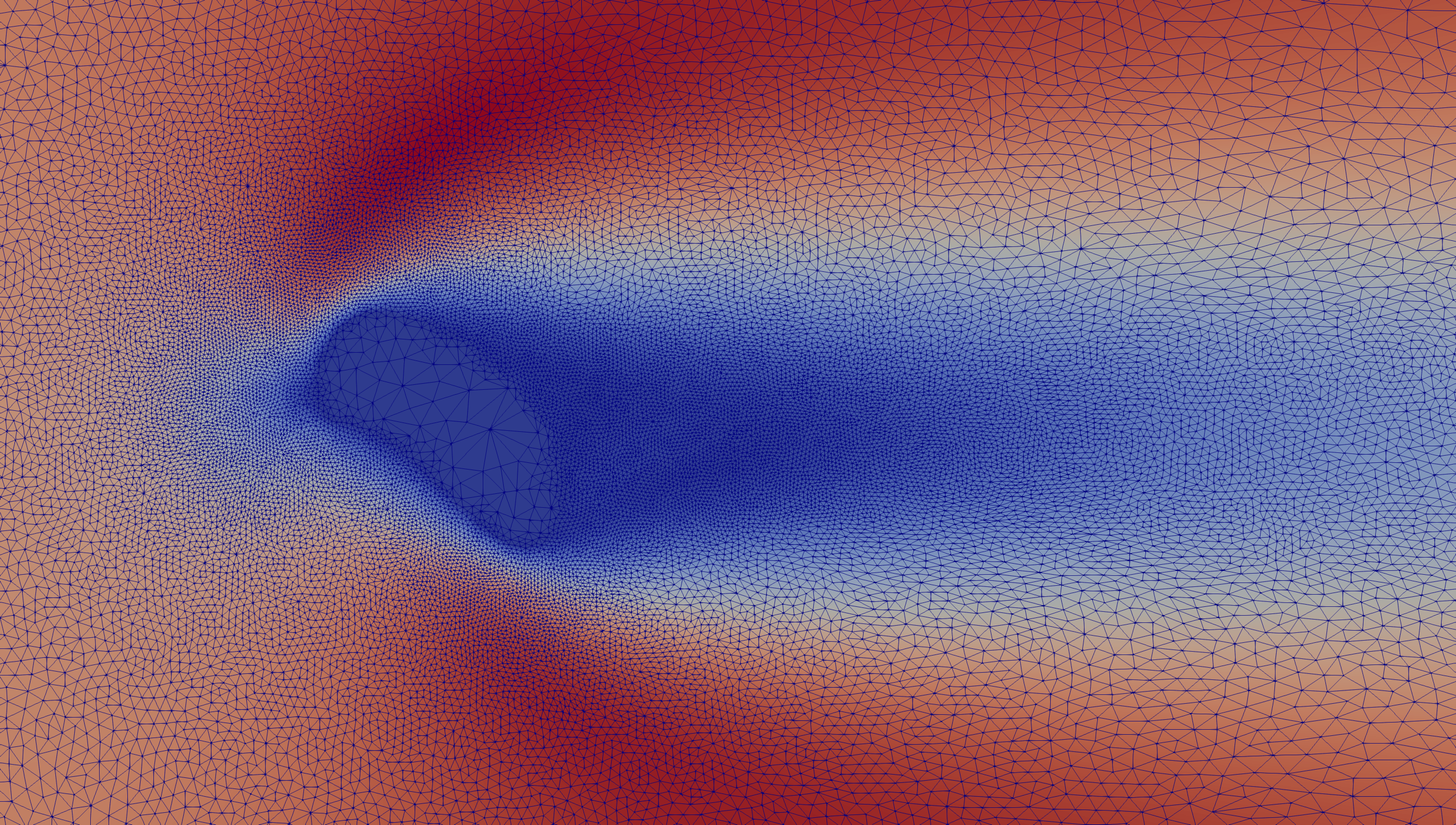}}
	
	\bigskip
	\bigskip
	\bigskip
	\smallskip
	\smallskip
	
	\caption{Final velocity field with mesh representation (the remaining of the domain is cropped)}
\end{subfigure}

\caption{\textbf{Results of a CFD computation.} \textit{Left}: The physical time is chosen large enough so the drag and lift coefficient values are stabilized. \textit{Right}: The remeshing technique increases the resolution (i) at the fluid/solid interfaces and (ii) in the areas presenting a high velocity gradient, thus ensuring accurate results with a limited computational charge.}
\label{fig:cfd}
\end{figure}

\noindent All the labels were computed using CimLib \cite{Hachem2013}, following the methods exposed in sections \ref{section:NS}, \ref{section:immersed} and \ref{section:equations}. \blue{The CFD solver used is equipped with a remeshing technique able to track both (i) the fluid/solid interfaces, and (ii) the areas of high velocity gradients. This method, exploited in conjunction with mesh immersion, ensures accurate results for the creation of the dataset (see figure \ref{fig:cfd}). Given the situation (multiple cheap 2D simulations), each CFD run was processed on a single core, with 64 shapes running at the same time on Intel Xeon 2.6 GHz cores. The average computation time was 4.8 minutes, and the whole dataset was generated in less than 24 hours. The physical computational time was chosen large enough so that the stationary flow was established, and that the computed drag and lift coefficients were stabilized (see figure \ref{fig:cfd}).}

\section{Results}

\subsection{Baseline network performance}
\label{section:baseline}

\begin{figure}
\centering

\setlength{\fboxsep}{0pt}%
\setlength{\fboxrule}{1pt}%

\begin{subfigure}[b]{.45\textwidth}
	\hspace{-18pt} \includegraphics{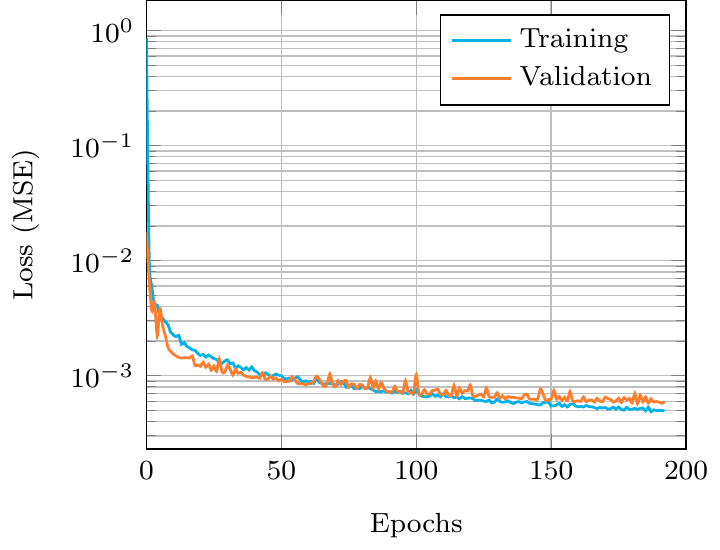}
	\caption{Training and validation loss}
	\label{fig:training}
\end{subfigure} \quad
\begin{subfigure}[b]{.45\textwidth}
	\hspace{-18pt} \includegraphics{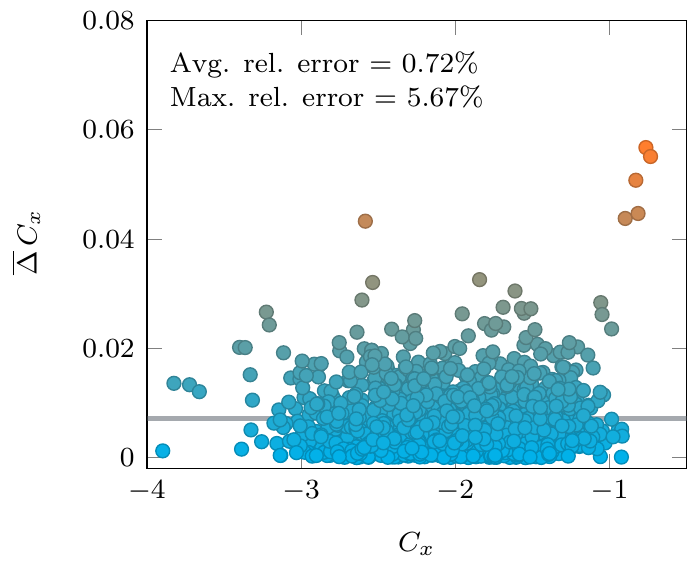}
	\caption{Baseline relative error on drag prediction}
	\label{fig:baseline_performance}
\end{subfigure}

\bigskip
\bigskip

\begin{subfigure}[b]{.45\textwidth}
	\centering
	\begin{subfigure}{.3\textwidth}
		\centering
		\fbox{\includegraphics[width=\textwidth]{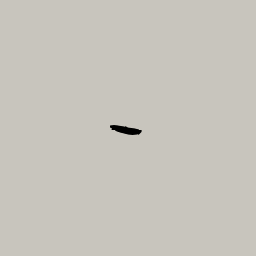}}
	\end{subfigure}
	\begin{subfigure}{.3\textwidth}
		\centering
		\fbox{\includegraphics[width=\textwidth]{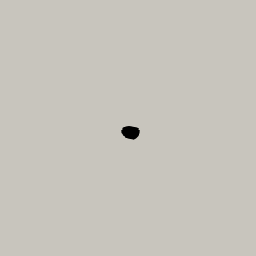}}
	\end{subfigure}
	\begin{subfigure}{.3\textwidth}
		\centering
		\fbox{\includegraphics[width=\textwidth]{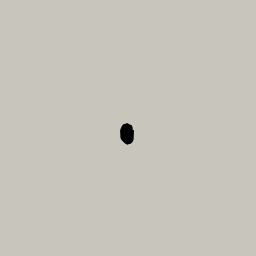}}
	\end{subfigure}
	
	\begin{subfigure}{.3\textwidth}
		\centering
		\fbox{\includegraphics[width=\textwidth]{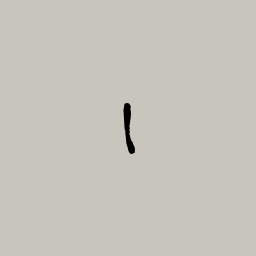}}
	\end{subfigure}
	\begin{subfigure}{.3\textwidth}
		\centering
		\fbox{\includegraphics[width=\textwidth]{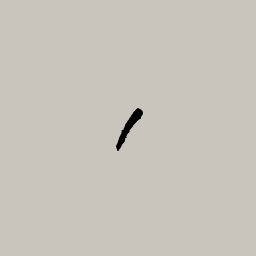}}
	\end{subfigure}
	\begin{subfigure}{.3\textwidth}
		\centering
		\fbox{\includegraphics[width=\textwidth]{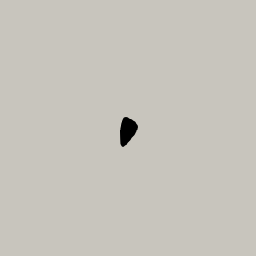}}
	\end{subfigure}

	\medskip
	\caption{Six worst performing shapes}
	\label{fig:baseline_shapes}
\end{subfigure}
\caption{\textbf{Results for the baseline network.} \textit{Top left}: the training and validation loss as a function of epochs. The early stopping technique helps avoiding overfitting of the network. \textit{Top right}: the average and maximal relative drag prediction errors over the test subset remain low, showing the good generalization capabilities of the network. \textit{Bottom}: the worst performing shapes are among those with smallest areas, as the input image definition is the same for all shapes.}
\label{fig:baseline}
\end{figure}

\blue{Here, we assess the performance of the baseline network introduced in section \ref{section:nn_imp} on the drag prediction task. As said earlier, the input images are of size $128 \times 128$, and the total number of learnable parameters is 66,497. In the remaining of the paper, the learning rate is set to \num{1e-3} with a decay factor of $\num{5e-3}$, and early stopping is used to determine the end of the training. The network parameters are initialized randomly, without pre-training, and a batch size of 64 is used by default. The training is processed on a Tesla V100 GPU card, on which one epoch requires approximately 5 seconds, for a total training time of 662 seconds. We use the Adam optimizer, with mean-squared error as loss. The training and validation loss curves are shown in figure \ref{fig:training}.} 

\blue{The predictive performance of the network is then computed by measuring the relative drag prediction error on the test subset. To do so, a forward network pass is made for each shape of the subset to obtain the predicted drag, which is compared to the exact drag. The relative prediction error is then computed. A plot of the error levels on the test set is shown in figure \ref{fig:baseline_performance}. The low average relative error indicates a good overall accuracy, except for some shapes presenting a low drag (roughly, $C_x \leq 1$), for which levels as high as 5\% can be reached. As could be expected, the worst performing shapes are that with the smallest areas (see figure \ref{fig:baseline_shapes}), as the input image resolution is the same for all shapes}.

%
%
%

\subsection{Batch size}

\begin{figure}
\centering

\begin{subfigure}[t]{.45\textwidth}
	\hspace{-18pt} \includegraphics{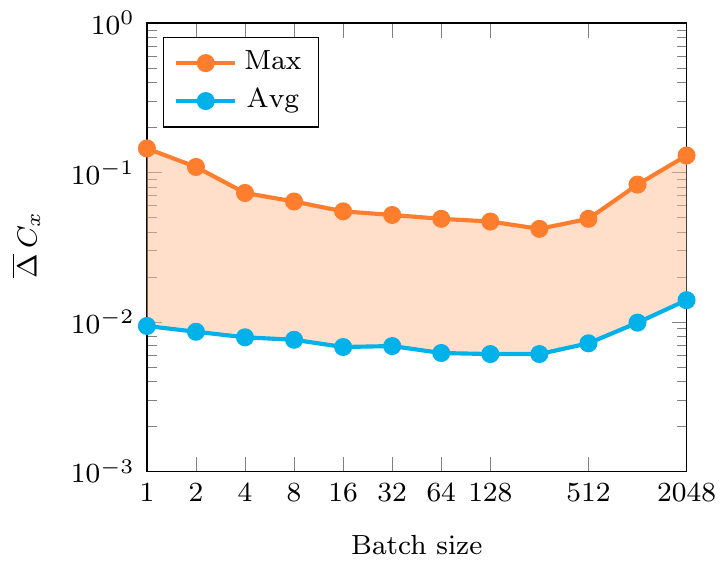}
	\caption{Average and maximal relative error obtained for drag prediction using different batch sizes.}
\end{subfigure} \quad
\begin{subfigure}[t]{.45\textwidth}
	\hspace{-18pt} \includegraphics{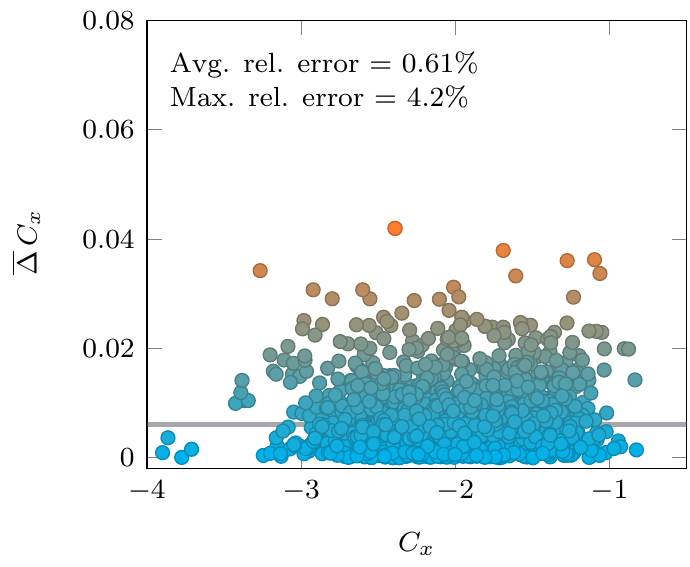}
	\caption{Relative error on drag prediction for a batch size equal to 256.}
\end{subfigure}
\caption{\textbf{Analysis of the influence of batch size on the average and maximal relative error.} A slight minimum is obtained using a batch size of 256, for both maximum and average drag relative errors.}
\label{fig:batch_size}
\end{figure}

The choice of the batch size in supervised learning is known to have a major impact on the performances of the resulting network \cite{Goodfellow2017}. It has multiple outcomes, such as (i) the accuracy of the gradient estimate, (ii) the time required for training or (iii) the necessity of an adequate learning rate. In figure, \ref{fig:batch_size}, we plot the average and maximal relative prediction errors as a function of the batch size using the baseline network of section \ref{section:baseline}. \blue{As in the previous cases, early stopping was used to prevent overfitting}. As can be seen, a slight minimum is obtained for both average and maximal relative errors when using a batch size of 256. The relative error over the test subset is also shown in figure \ref{fig:batch_size}: as can be seen, the high prediction errors obtained for the smaller-sized shapes in figure \ref{fig:baseline_performance} have now dropped down to a level similar to that of other shapes of larger sizes, with a maximal error level as low as 4.2\%. \blue{In the following, the batch size is set equal to 256}.

\subsection{Network optimization}

\begin{figure}
\centering
\includegraphics{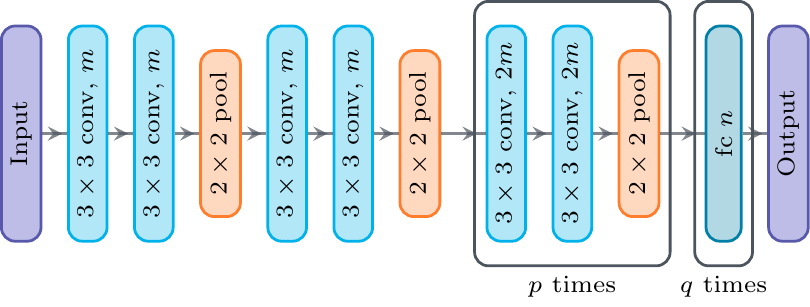}

\caption{\textbf{Network pattern for optimization.} The goal is to optimize separately the convolutional part ($m$ and $p$ parameters) and the fully connected part ($n$ and $q$ parameters).}
\label{fig:arch_opt}
\end{figure}

\blue{In this section, we optimize the network architecture further by modifying both convolutional and fully-connected parts. To do so, we consider the generic architecture shown in figure \ref{fig:arch_opt}. This architecture is a generalization of the baseline one, which was obtained by trial and error. The goal is to explore variants of this baseline network version, by varying separately the depth and complexity of (i) the convolutional part ($m$ and $p$ parameters), and (ii) the fully-connected part ($n$ and $q$ parameters).}

\blue{The convolutional and fully connected parts of the network are optimized separately: first, the convolutional part is considered, by varying $m$ in $\left[1,64\right]$ and $p$ in $\left[0,4\right]$, while keeping the fully-connected block of the baseline network. For each case, the maximal and average relative errors of the network on the test subset are computed, and shown in figures \ref{fig:conv_opt_max} and \ref{fig:conv_opt_avg}. Increasing the amount of filters per layer $m$ is clearly beneficial for any value of $p$. The effect of increasing network depth $p$ is not as clear, although the best configurations are obtained for $p=3$ or $p=4$. As optimal performance for average and maximum relative errors are not obtained for the same $(p,m)$ couples, small network sizes are favoured. For that reason, we choose $(p_{\text{opt}},m_{\text{opt}}) = (3,32)$.}

\begin{figure}
\centering

\begin{subfigure}[b]{.45\textwidth}
	\centering
	\includegraphics{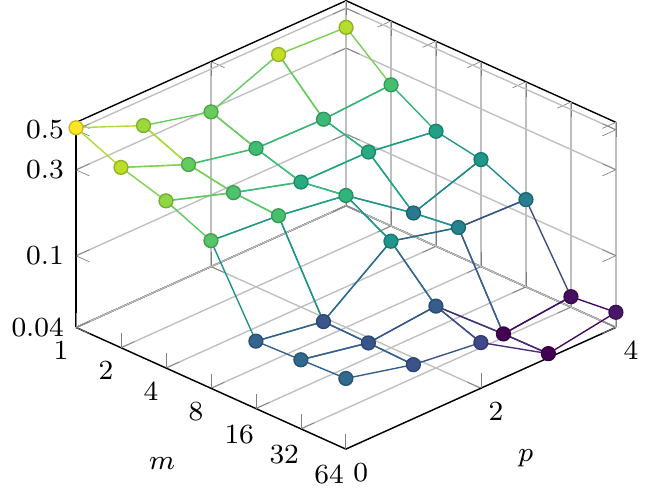}
	\caption{Maximal drag relative error over test subset as a function of $p$ and $m$}
	\label{fig:conv_opt_max}
\end{subfigure} \quad \quad
\begin{subfigure}[b]{.45\textwidth}
	\centering
	\includegraphics{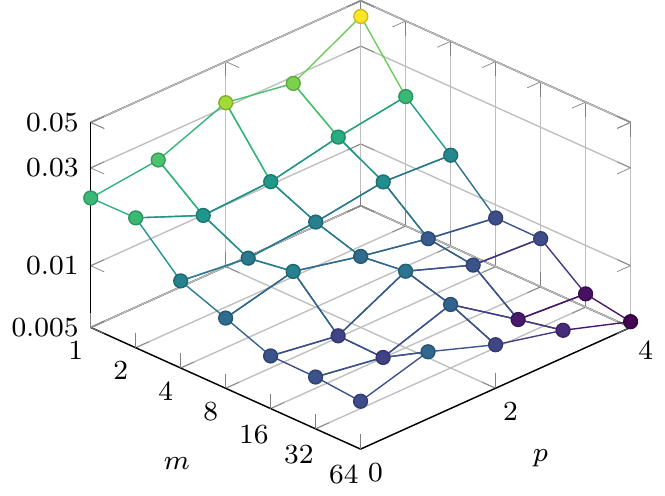}
	\caption{Average drag relative error over test subset as a function of $p$ and $m$}
	\label{fig:conv_opt_avg}
\end{subfigure}

\caption{\textbf{Optimization of the convolutional block of the network.} Both plots represent the drag relative error as a function of $p$ and $m$ (see figure \ref{fig:arch_opt}). \textit{Left:} maximal relative error. \textit{Right:} average relative error.}
\label{fig:conv_opt}
\end{figure}

\begin{figure}
\centering

\setlength{\fboxsep}{0pt}%
\setlength{\fboxrule}{1pt}%

\begin{subfigure}[b]{.45\textwidth}
	\centering
	\includegraphics{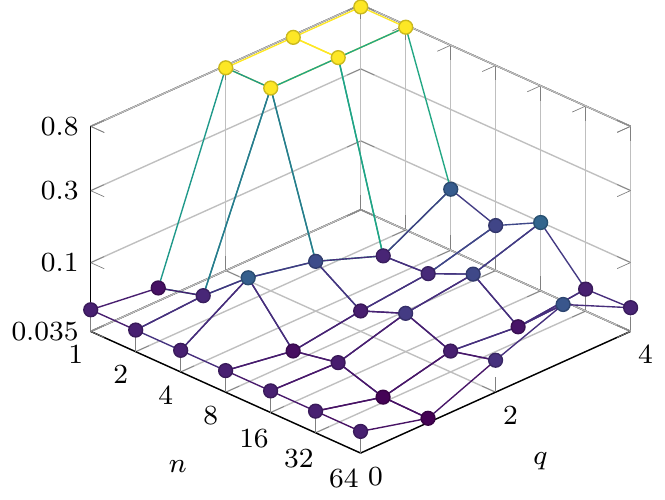}
	\caption{Maximal drag relative error over test subset as a function of $q$ and $n$}
	\label{fig:conv_opt_max}
\end{subfigure} \quad \quad
\begin{subfigure}[b]{.45\textwidth}
	\centering
	\includegraphics{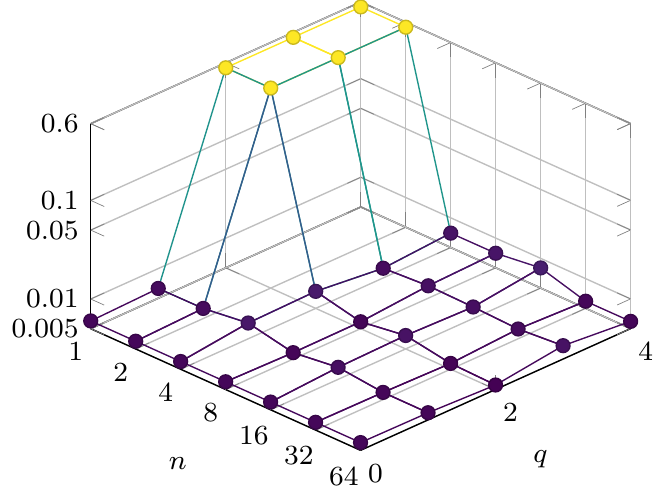}
	\caption{Average drag relative error over test subset as a function of $q$ and $n$}
	\label{fig:conv_opt_avg}
\end{subfigure}

\caption{\textbf{Optimization of the fully connected block of the network.} Both plots represent the drag relative error as a function of $q$ and $n$ (see figure \ref{fig:arch_opt}). \textit{Left:} maximal relative error. \textit{Right:} average relative error.}
\label{fig:fc_opt}
\end{figure}

\begin{figure}
\centering
\includegraphics{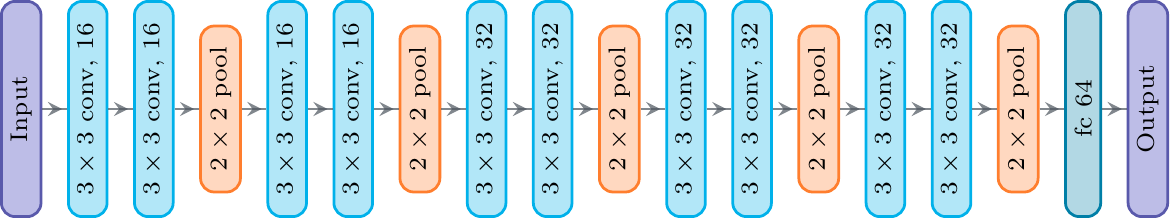}

\caption{\textbf{Optimized drag prediction CNN.} This architecture was obtained by optimizing the baseline network through a hyper-parameter search.}
\label{fig:net_opt}
\end{figure}

\blue{In CNNs, the task attributed to fully-connected layers is to learn non-linear combinations of the high-level features extracted by the convolutional blocks. Modern classification networks such as VGG \cite{Simonyan2014} or ResNet \cite{He2016} usually include zero to a few dense layers between the convolutional blocks and the output layer. Here, $n$ varies in $\left[1,64\right]$, while $q$ varies in $\left[0,4\right]$: results are shown in figure \ref{fig:fc_opt}. Interestingly, very decent performance is obtained when the output of the last convolutional layer is flattened and directly fed to the output layer (\ie for $q=0$). Another noticeable point is that using several fully connected layers with very few neurons per layer (\ie $q \geq 2$ and $n \leq 2$) leads to almost no learning. As for the optimization of the convolutional block, minimal errors are obtained on different configurations for maximum and average errors. We follow the same line as before by choosing the configuration leading to the smallest network, and therefore we set $(q_{\text{opt}},n_{\text{opt}}) = (1,64)$. Eventually, the best network obtained, shown in figure \ref{fig:net_opt}, holds 296,865 learnable parameters. Each training epoch requires approximately 6 seconds, leading to a total training time of of 1535 seconds.}

\subsection{Drag prediction on realistic shapes}

We now evaluate the predictive capabilities of the best network on a selected set of shapes, including geometrical shapes (cylinder, square) and NACA airfoils. The results are summed up in table \ref{table:shapes_predictions}. The shapes dimensions are adapted to fit the mean dimensions of the dataset shapes, \ie they fit in the $\left[-1,1\right]^2$ square, with their center of mass centered in (0,0). Relative error levels remain low on such shapes, with a maximal value of 3.06\% for the horizontal bar. As could be expected, the prediction on a shape of the dataset yields a very accurate result, with a relative error below 0.2\%. Finally, the error levels for NACA airfoils remain low, with a maximum level of 1.27\%. This experiment also underlines the interest of the random shape dataset for the drag prediction on non-random, real-life shapes.

\newcolumntype{C}[1]{ >{\vbox to 4ex\bgroup\vfill\centering\arraybackslash}p{#1}<{\vskip-\baselineskip\vfill\egroup}}  

\begin{table}
\footnotesize
\caption{\textbf{Exact and predicted drags for several handpicked shapes.} The shapes largest dimensions were adapted to fit the mean dimensions of the dataset shapes. The different geometrical parameters given in the array are the following: $w$ stands for \emph{width}, $h$ stands for \emph{height}, $r$ stands for \emph{radius} and $c$ stands for \emph{chord}. It is important to notice that in the following table, the scale of the NACA airfoils is not that of the other shapes.}
\label{table:shapes_predictions}
\medskip
\centering
\begin{tabular}{C{3.5cm}C{3.5cm}C{2cm}C{1.5cm}}
\toprule
Shape 										& Description					& Prediction (rel. error)  	& Exact drag 	\\\midrule
\includegraphics[width=.05\linewidth]{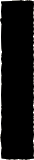}		& Vertical bar, $h=1$, $w=0.2$		& 1.585 (0.69\%)		& 1.596		\\\midrule
\includegraphics[width=.25\linewidth]{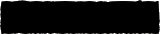}	& Horizontal bar, $h=0.2$, $w=1$	& 0.978 (3.06\%)		& 0.949		\\\midrule
\includegraphics[width=.25\linewidth]{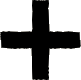}		& Cross, $w=1$, $h=0.2$			& 1.571 (1.16\%)		& 1.553		\\\midrule
\includegraphics[width=.2\linewidth]{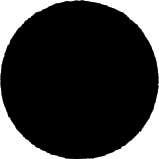}		& Cylinder, $r=0.5$				& 1.586 (0.19\%)		& 1.589		\\\midrule
\includegraphics[width=.2\linewidth]{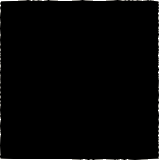}		& Square, $h=w=1$				& 1.763 (0.056\%)		& 1.764		\\\midrule
\includegraphics[width=.2\linewidth]{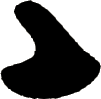}		& Random shape from DS			& 1.894 (0.16\%)		& 1.897		\\\midrule
\includegraphics[width=.6\linewidth]{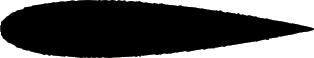} 	& NACA 0018, $c=1$			& 1.192 (0.51\%)		& 1.186 		\\\midrule
\includegraphics[width=.6\linewidth]{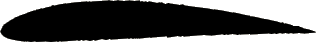}	& NACA 4412, $c=1$			& 1.111 (0.89\%)		& 1.121		\\\midrule
\includegraphics[width=.6\linewidth]{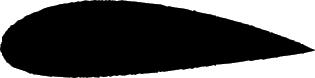} 	& NACA 4424, $c=1$			& 1.279 (1.27\%)		& 1.263 		\\\midrule
\includegraphics[width=.6\linewidth]{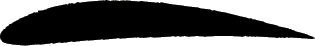}	& NACA 6412, $c=1$			& 1.119 (1.15\%)		& 1.132		\\\bottomrule
\end{tabular}
\end{table}


\subsection{Conclusion}

In this paper, an optimized convolutional neural network was introduced for the drag prediction of arbitrary 2D shapes in laminar flow at $Re=10$. This network was trained on a custom dataset composed of 12,000 random shapes built with B\'ezier curves, and which drag was computed numerically by solving the Navier-Stokes equations using an immersed mesh method. The large variety of geometrical shapes in the dataset allowed the network to make accurate drag predictions on realistic shapes such as NACA airfoils, with a maximal relative error in the 1-2\% range.

These results underline the potential of this approach, and shall be pursued at higher Reynolds. \blue{In the case of turbulent flows, the prediction of a time-averaged quantity of interest can be considered (see \cite{Miyanawala2017}). Still, in many industrial-level CFD computations, a RANS turbulence model is considered that leads to converged, stationary drag and lift values even at high Reynolds numbers.} \blue{The extension of the current work to three-dimensional shapes can also be considered, using 3D CNNs.} Also, exploiting advanced network architectures, such as ResNets or densely connected convolutional networks, may lead to even better results.

\newpage
\bibliographystyle{unsrt}
\bibliography{nn_flow_2d}

\end{document}